# DeepDoseNet: A Deep Learning model for 3D Dose Prediction in Radiation Therapy

Mumtaz Hussain Soomro[1], Victor Gabriel Leandro Alves[1], Hamidreza Nourzadeh[1,2], Jeffrey V. Siebers[1]

[1]University of Virginia Health System, Charlottesville, VA

[2]Thomas Jefferson University Hospital, Philadelphia, PA

**Abstract:** **Purpose:** To describe an artificial intelligence — deep learning-based method for fully automated, accurate, and rapid prediction of optimized head-and-neck 3D dose distributions for the American Association of Physicists (AAPM) in Medicine Open Knowledge-Base Planning Challenge (OpenKBP), and inter-compare the effect of alternative loss functions on the method's performance.

**Method:** The DeepDoseNet 3D dose prediction model is based on ResNet and Dilated DenseNet. The 340 head-and-neck datasets from the 2020 AAPM OpenKBP challenge were utilized, with 200 for training, 40 for validation, and 100 for testing. Structures include 56Gy, 63Gy, 70Gy PTVs, and brainstem, spinal cord, right parotid, left parotid, larynx, esophagus, and mandible OARs. Mean squared error (MSE) loss, mean absolute error (MAE) loss, and MAE plus dose-volume histogram (DVH)-based loss functions were investigated. Each model's performance was compared using a 3D dose score, $\overline{S_D}$, (mean absolute difference between ground truth and predicted 3D dose distributions) and a DVH score, $\overline{S_{DVH}}$ (mean absolute difference between ground truth and predicted dose-volume metrics). Furthermore, DVH metrics Mean[Gy] and D0.1cc [Gy] for OARs and D99%, D95%, D1% for PTVs were computed.

**Results:** DeepDoseNet with the MAE plus DVH-based loss function had the best dose score performance of the OpenKBP entries. On the models we tested, the MAE+DVH model had the lowest prediction error ($P<0.0001$, Wilcoxon test) on validation and test datasets (validation: $\overline{S_D}$=2.3Gy, $\overline{S_{DVH}}$=1.9Gy; test: $\overline{S_D}$=2.0Gy, $\overline{S_{DVH}}$=1.6Gy) followed by the MAE model (validation: $\overline{S_D}$=3.6Gy, $\overline{S_{DVH}}$=2.4Gy; test: $\overline{S_D}$=3.5Gy, $\overline{S_{DVH}}$=2.3Gy). The MSE model had the highest prediction error (validation: $\overline{S_D}$=3.7Gy, $\overline{S_{DVH}}$=3.2Gy; test: $\overline{S_D}$=3.6Gy, $\overline{S_{DVH}}$=3.0Gy). No significant difference was found among models in terms of Mean [Gy], but the MAE+DVH model significantly outperformed the MAE and MSE models in terms of D0.1cc[Gy], particularly for mandible and parotids on both validation (P < 0.01) and test (P < 0.0001) datasets. MAE+DVH outperformed (*P*<0.0001) in terms of D99%, D95%, D1% for targets.

**Conclusion:** The use of the DVH-based loss function reduced $\overline{S_D}$ by ~60% and $\overline{S_{DVH}}$ by ~70%, and resulted in the best performance amongst OpenKBP entries. DeepDoseNet should be suitable for automated radiation therapy workflows, including clinical planning and studies of the dosimetric relevance of delineated organs-at-risk.

## Introduction

Current advancements in radiation therapy (RT) treatment planning are progressively improving cancer patients' treatment outcomes. In contrast to conventional 3D conformal therapy, modern treatment methods, such as intensity modulation radiation therapy (IMRT) [1] and volumetric arc therapy (VMAT) [2], have been practiced, delivering high therapeutic radiation dosage to the target while paring down the doses to surrounding organs-at-risk (OARs). The creation of an IMRT or VMAT treatment plan involves finding the optimal balance between competing clinical objectives [3], [4]. The inherently complex nature of these methods results in lengthy planning time and results in high variability in the treatment plan quality according to the planner's expertise and experience. To reduce the effort, several dose prediction methods [5]–[10], [11] based on knowledge-based planning (KBP) have been accomplished to improve treatment plan quality and efficacy, while decreasing plan variability by providing consistency and reducing human planning time efforts. A KBP-based commercial software, so-called RapidPlan (version 13.6, Varian Oncology Systems, Palo Alto, CA, USA) is currently available to estimate an optimal treatment plan [12].

However, the accuracy of current KBP-based dose prediction approaches is limited due to the challenges of including patient-specific characteristics and the fact that they primarily aim to reproduce the one-dimensional dose-volume histogram (DVH). The DVH alone insufficiently characterizes the 3D dose distribution, limiting the ability of current KBP methods to estimate the optimal 3D dose distribution [13] [14].

Recently, deep learning-based convolutional neural networks (CNNs) methods have been successfully applied with impressive results for medical image segmentation [15]–[24], object detection [25], [26], image registration [27], [28], and so on. Deep learning-based approaches learn features for intricate patterns and structures from well-organized large training datasets directly as a trained model of prior information. Subsequently, the trained model is applied to new unseen data for prediction without manually extracting the complex features. Deep learning CNNs circumvent the need for handcrafted features since CNN learns the relevant complex features needed for prediction. The success of deep learning-based approaches has motivated the deployment of CNNs for 3D voxel-wise dose distribution prediction using anatomical structures and/or computed tomography (CT) as an input, [29], [30], [31]–[34]. To resolve the challenges of an intercomparison of alternative dose-prediction models, in 2020, the AAPM organized the Open Knowledge-Based Planning Challenge (OpenKBP), which provided common open-source data for all participants pre-identified standardized comparison metrics [35].

This work describes our OpenKBP model, a 3D CNN model called DeepDoseNet based on Residual networks [36] and DenseNet [37]. DeepDoseNet predicts 3D voxel-wise dose distributions for head and neck cancer IMRT patients. Three different loss functions (mean square error (MSE), mean absolute error (MAE), and MAE plus DVH) are inter-compared with DeepDoseNet to explore the loss-function's effect on the model's performance.

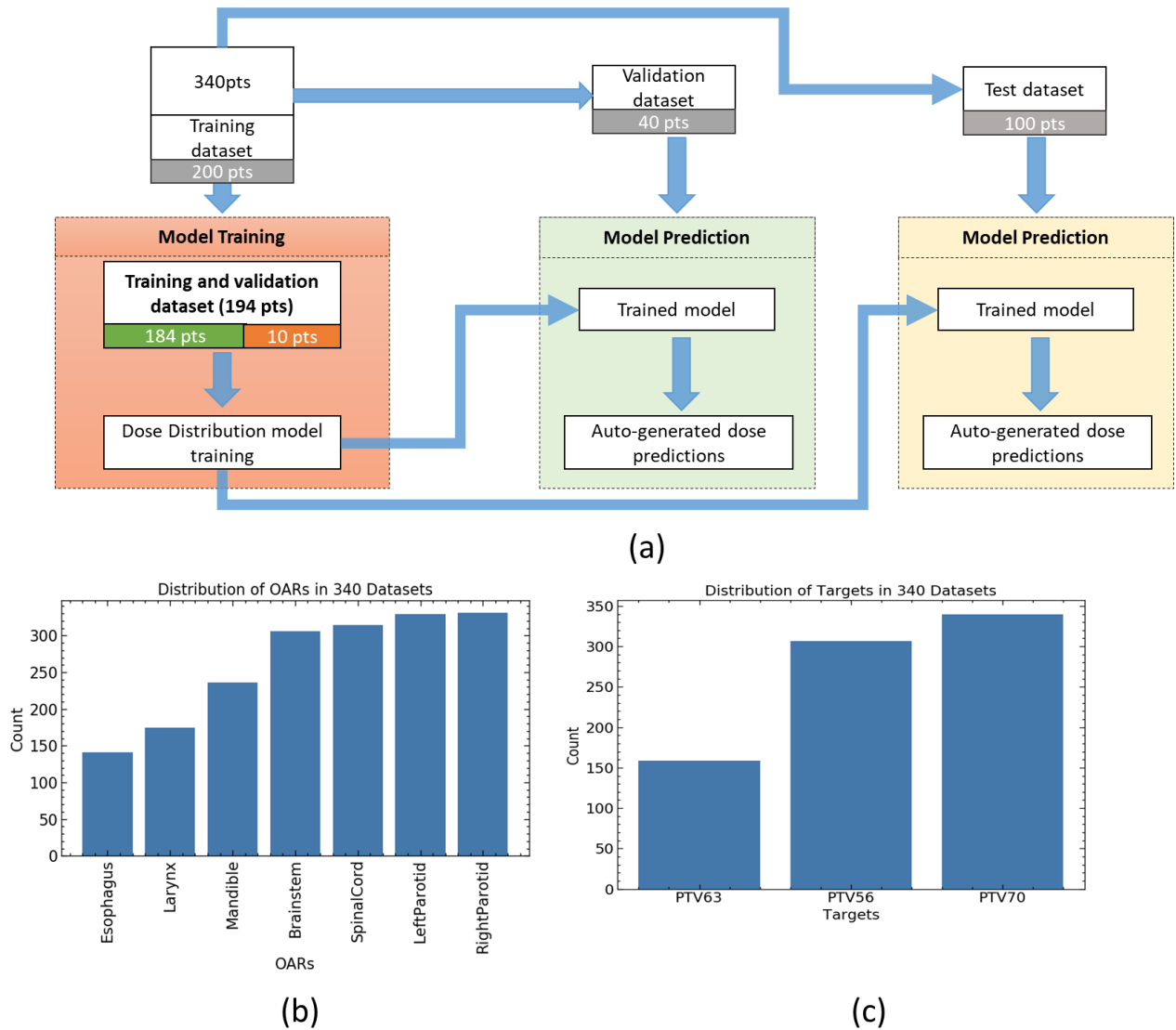

Figure 1: (a) Flowchart is showing the division of data into training, validation, and testing datasets; (b) distribution of organs-at-risk (OARs) in datasets; (c) distribution of targets (PTV56, PTV63, PTV70) in the datasets.

## Materials and Methods

### Datasets

This study used the 340 Head and Neck patient dataset prepared by OpenKBP [35]. OpenKBP obtained the CT structure datasets from The Cancer Imaging Archive (TCIA) [38], which OpenKBP standardized in terms of structure format, naming, and planning technique. In addition to down-sampling each CT to a 128×128×128 voxel tensor, OpenKBP provided structure masks for targets and OARs used in the planning, as well as the plan dose distributions. The distribution of OARs and targets is shown in Figure 1 (b) and (c). The reference treatment plans were generated with CERR [39], yielding reference dose distributions [40] with similar fluence complexity [41]. Each plan utilized nine equispaced coplanar 6 MV step-and-shoot intensity-modulated radiation therapy (IMRT) beams optimized to yield planning target volume (PTV) prescription doses of 56Gy, 63Gy, 70Gy in 35 fractions while limiting the dose to seven OARs (brainstem, spinal cord, right parotid, left parotid, larynx, esophagus, and mandible).

The OpenKBP challenge specified that 200 datasets be used for training, 40 for validation, and 100 for testing. An overview of the data division is depicted in Figure 1 (a).

### DeepDoseNet

The overall network architecture used to predict the dose distributions is depicted in Figure 2 (a). The architecture is adopted from 3D Unet [18], which follows the classical encoder-decoder type network architecture. Unlike a standard 3D Unet, we incorporated residual network (Resnet) [36] blocks in the encoder and decoder instead of a standard convolution layer stack at each resolution stage. Each Resnet block consists of two $3\times3\times3$ sized kernel convolution layers, where each convolution layer is followed by batch normalization and ReLU [42]. Inspired from [43], [17], and [44], our network includes dense multiscale feature aggression (MFA) between the encoder and decoder, similar to [45]. The encoder comprises of a convolution layer with kernel size = $3\times3\times3$, stride = 2, three ResNet blocks, where a $2\times2\times2$ max-pooling layer follows each ResNet block. The primary purpose of the MFA is to extract multiscale features with enlarging receptive fields without downgrading the features' resolution. MFA utilizes dilated convolutions that have attained an effective state-of-art performance in semantic segmentation tasks [43], [46], and object detection tasks [47]. The dilated convolution effectively increases the receptive field without employing a pooling (i.e., down-sampling) operation and has an equal number of parameters as an ordinary convolution layer [43]. This type of mechanism preserves the dose distribution details for small-sized organs (i.e., brain stem or spinal cord) by providing a large receptive field for them. Inspired by the previous study of dense feature extraction in semantic segmentation [44], MFA is designed on a dilated DenseNet block. The dilated DenseNet block contains five consecutive $3\times3\times3$ sized kernel convolution layers, where the dilation rate in each convolution layer is set as 1, 2, 5, and 9 that varies their corresponding receptive field as $3\times3\times3$, $5\times5\times5$, $11\times11\times11$, and $19\times19\times19$, respectively.

The network layers are densely connected by following dense connections as in [37]; the dense connectivity reduces the number of training parameters compared with other networks [37], [48]. The decoder part gradually recovers the low-resolution features of MFA and produces the final high-resolution output (i.e., predicted dose). The decoder part consists of four 2×2×2 up-sampling (transpose convolution) with stride = 2, three Resnet blocks, and a final 1×1×1 convolution layer followed by ReLU to generate the predicted dose map.

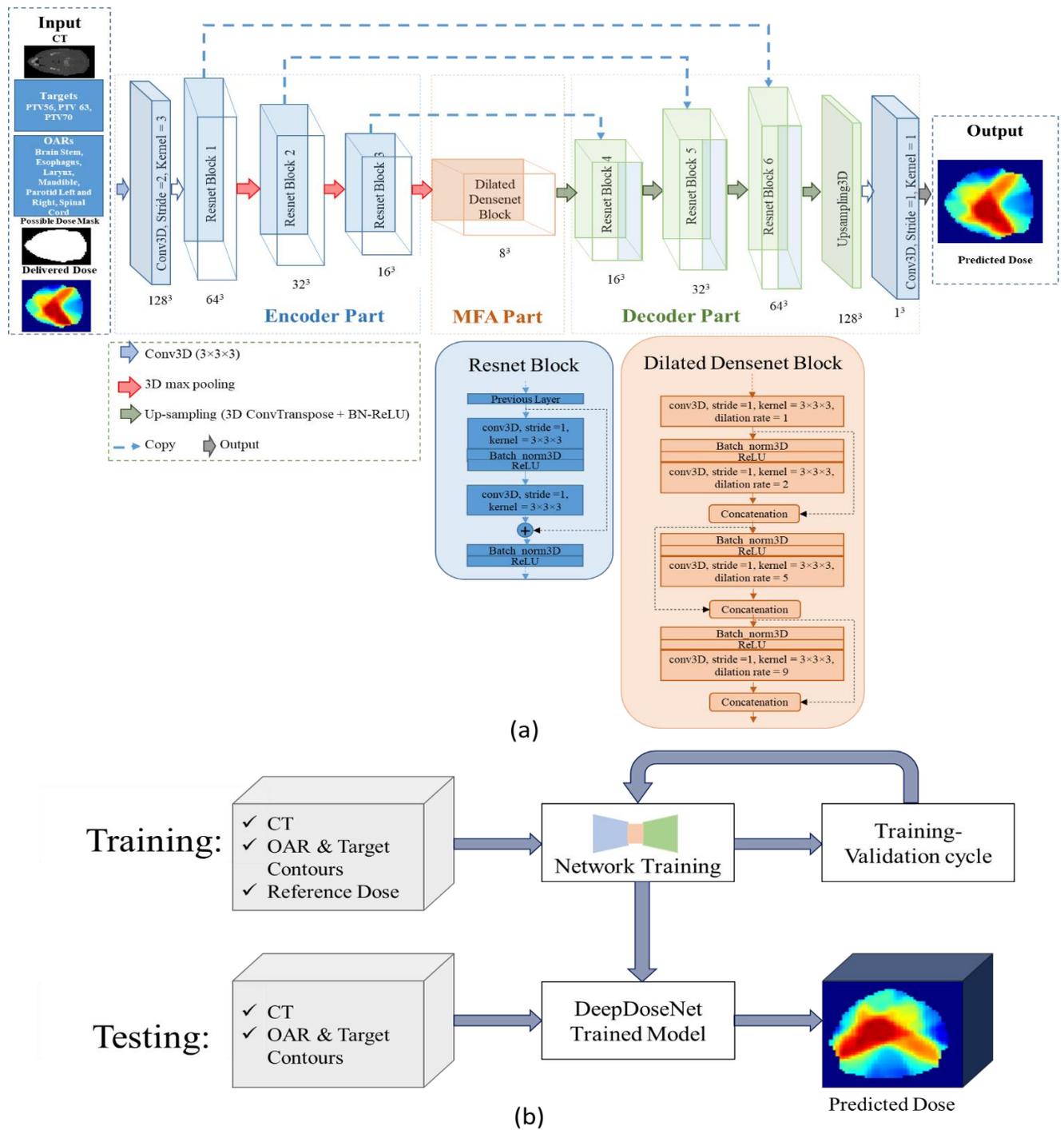

Figure 2: a) The proposed DeepDoseNet architecture; (b) Flowchart representing training and testing (prediction) phases of DeepDoseNet.

DeepDoseNet – Network training

The DeepDoseNet model's inputs vary from 4 to 12 channels depending on the number of OARs and targets contours in each dataset (i.e., input shape = 128×128×128×12) and its output (predicted dose) with a shape of 128×128×128×1. The OpenKBP provided dose distribution was utilized as the reference (ground truth). Models were developed and intercompared based on three different loss functions during training; *MSE*, *MAE*, and *DVH*. For each model, the Adam optimizer [49] with learning rate = 0.001, decay = 0.0001, β1 = 0.9, β2 = 0.99 (momentum) was used to minimize the loss function between the predicted dose and the ground truth. Each model was trained using a hold-out method with 200 epochs with batch size = 2. The training was conducted in PyTorch [50] on an Nvidia Quadro 48000 GPU with 48 GB memory. No data augmentation strategy was applied, as the authors believe dose distributions follow certain physics constraints, such as maximum achievable dose gradients (i.e., Gy/mm), beam arrangements, and radiation field orientation, which are created at the treatment machine's coordinate system. Augmentations via, e.g., applying elastic transformations on the dose dataset would violate the physics constraints, hence, were avoided. After the models were fully trained, 3D dose predictions on new unseen data are generated in ~0.5 seconds per dataset. An overview of training and testing (prediction) phases DeepDoseNet is illustrated in Figure 2 (b).

## Loss Functions

Considering domain knowledge in Radiation Oncology, typical IMRT dose distributions have a high target (PTV) doses surrounded by steep dose gradients to spare surrounding OARs. This physical property is essential to the approach used when seeking deep learning-based solutions. We intercompared the ability of a DNN trained with different loss functions to capture these properties.

The mean square error (*MSE*) is the most common loss estimator used in machine learning. *MSE* is calculated as the average of the squared differences between the predicted and reference doses. Given a binary mask $M_d$ where the dose is computed, the loss function

$$MSE = \frac{\sum_{i,j,k}\left[\left(D_{pred}(i,j,k) - D_{ref}(i,j,k)\right)\right]^2 M_d(i,j,k)}{\sum_{i,j,k} M_d(i,j,k)}$$

where $D_{pred}(i,j,k)$ is the predicted dose for voxel $i,j,k$ and $D_{ref}(i,j,k)$ is the reference dose.

The *MSE* least-squares estimator is known to be highly sensitive to large deviations; as the difference is squared, prediction compromises required to reduce outliers can degrade the overall dose estimate. With IMRT dose gradients of ~15%/mm, accommodating dose-prediction results would be offset by a single 2 mm dose-voxel that can compromise the low-gradient homogeneous regions of dose distribution.

Robust Optimization (RO) using loss functions based on absolute deviations can overcome the *MSE* limitations [51]–[54] and are expected to improve the optimization result while capturing the physical properties of high-dose gradients between PTVs and OARs [55], [56]. As such, we implemented a mean absolute error (*MAE*) loss function, which is the average of the absolute differences between the predicted and reference doses given $M_d$:

$$MAE = \frac{\sum_{i,j,k} |(D_{pred}(i,j,k) - D_{true}(i,j,k))| M_d(i,j,k)}{\sum_{i,j,k} M_d(i,j,k)}.$$

Given a large number of voxels in dose distribution, *MAE* approximates the Nonlinear Least Absolute Deviations robust estimator proposed by Hitomi et al. [52], which converges to a uniquely minimized function.

While *MAE* improves agreement for high dose gradient regions, it alone does not map the fundamental preference in radiotherapy to treat/spare specific PTVs/OARs as opposed to unspecified tissue. In treatment plan optimization, this preference is often embodied in the dose and dose-volume objectives and constraints. Expecting that incorporation of this domain knowledge into dose prediction would enhance its performance, we added a DVH loss function to our model. To efficiently implement the DVH loss function, we utilized the differential approximation of the DVH $(\widetilde{DVH})$ by Nguyen *et al.*,[57]. Knowing that a DVH computation can be affected by voxel size/rounding approximations, similar to the *MSE* and *MAE*, the DVH loss function utilized the minimum absolute deviation instead of a mean-squared loss:

$$DVH(D_{true}, D_{pred}, M) = \frac{1}{n_s} \frac{1}{n_t} \sum_s |\widetilde{DVH}_s(D_{true}, M_s) - \widetilde{DVH}_s(D_{pred}, M_s)|,$$

where $n_s$ is the number of patient OAR structures, $n_t$ is the number of dose bins used to compute the DVH, and $M_s$ represents the binary segmentation masks for a given OAR.

Since the number of OARs differs per patient, the training dataset class imbalance is compensated by having the DVHs restricted to the available OARs per patient. DVH loss is implemented in a hybrid loss function combining MAE as the first-order metric and the DVH loss (i.e., MAE+DVH) as a regularization parameter.

## Performance Metrics

DeepDoseNet was evaluated with the 3D dose and DVH scores metrics utilized by the OpenKBP challenge for test and validation datasets[35]. The 3D dose score is the voxel-by-voxel mean absolute difference of the predicted volumetric dose within a mask $M_d$, and the DVH score is the mean absolute difference of DVH metrics including Mean[Gy], D0.1cc[Gy], D1%[Gy], D95%[Gy], and D99%[Gy]). The lowest value of the 3D dose score and DVH score indicates the model's best performance. We additionally report related DVH metrics; Mean[Gy] and D0.1cc[Gy] for OARs and D1%[Gy], D95%[Gy], and D99%[Gy] for targets (i.e., PTV56, PTV63, and PTV 70 structures).

## Results

### Learning curves

Figure 3 shows sample learning curves for the MSE, MAE, MAE+DVH models. Each model's validation loss decreases along with a decrement in training loss, revealing that each model does not exhibit serious overfitting. MAE yields a lower value of the loss function than MSE on training and validation losses. MAE+DVH loss function is more stable than MSE or MAE alone.

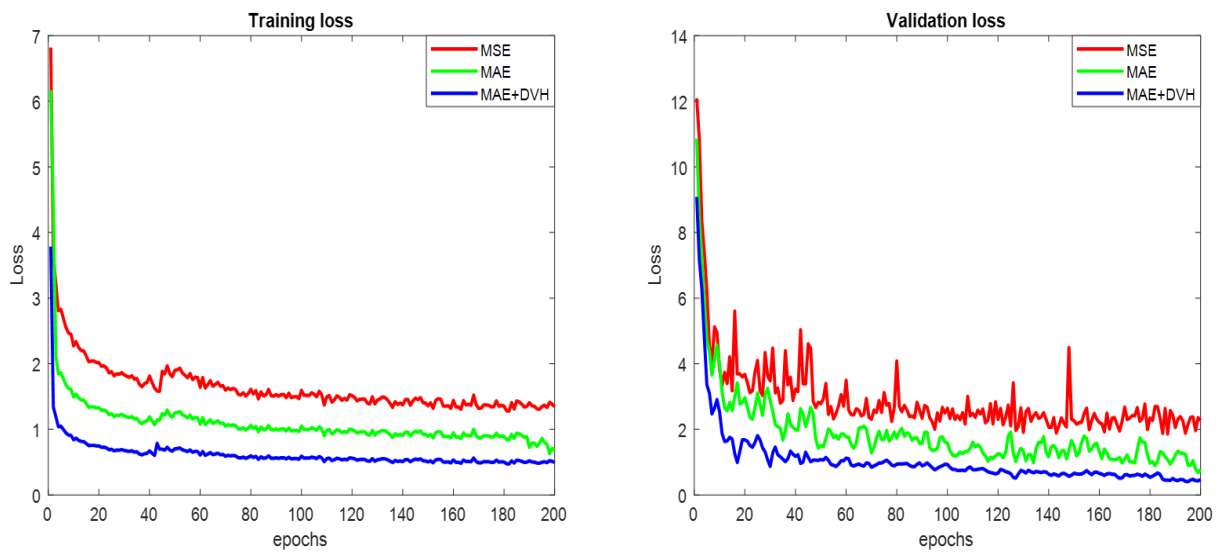

Figure 3: Training (left) and validation (right) losses of a model with different loss functions (MSE, MAE, MAE+DVH).

## Comparing Statistics of 3D Dose and DVH Scores

Figure 4 and Table 1 quantitatively compare the performance of MSE, MAE, MAE+DVH trained models in terms of DVH and 3D dose scores. There is no significant difference between MAE and MSE models on validation datasets for 3D dose score, but MAE outperforms MSE on test datasets ($P < 0.05$). In terms of DVH score, MAE model outperforms than MSE on both validation ($P < 0.001$) and test ($P < 0.0001$) datasets. The MAE+DVH model significantly outperformed both MSE and MAE models with the lowest 3D dose and DVH scores on validation and test datasets ($P < 0.0001$).

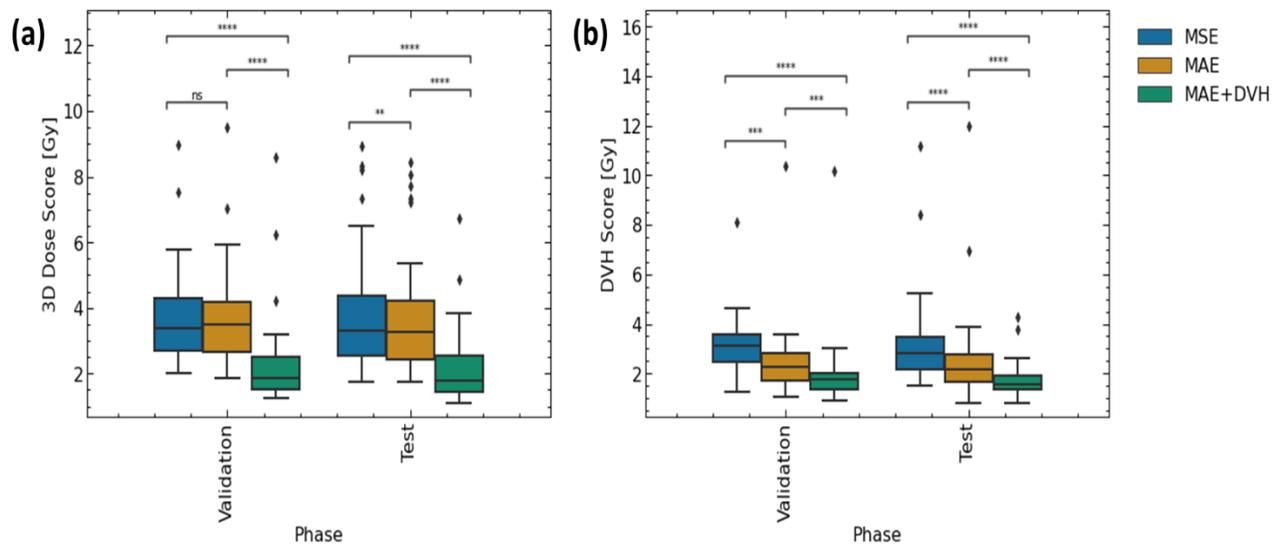

Figure 4 (a) Average 3D dose scores (mean absolute difference between predictions and reference dose distributions) and (b) average DVH scores differences on validation and test datasets. Wilcoxon test was used for analysis. Star (*) values shows different $P$ values, i.e., *$P< 0.05$, **$P< 0.01$, ***$P< 0.001$, ****$P< 0.0001$, ns = no significance.

Table 1: Average absolute Dose and DVH scores of the DeepDoseNet models (MSE, MAE, MAE+DVH) on validation and test datasets.

| | DeepDoseNet | | | |
|---|---|---|---|---|
| Loss function | phase | DVH Score [Gy] | 3D Dose Score [Gy] | N |
| MSE | validation | 3.2 | 3.7 | 40 |
| MSE | test | 3.0 | 3.6 | 100 |
| MAE | validation | 2.4 | 3.6 | 40 |
| MAE | test | 2.3 | 3.5 | 100 |
| MAE+DVH | validation | 1.9 | 2.3 | 40 |
| MAE+DVH | test | 1.6 | 2.0 | 100 |

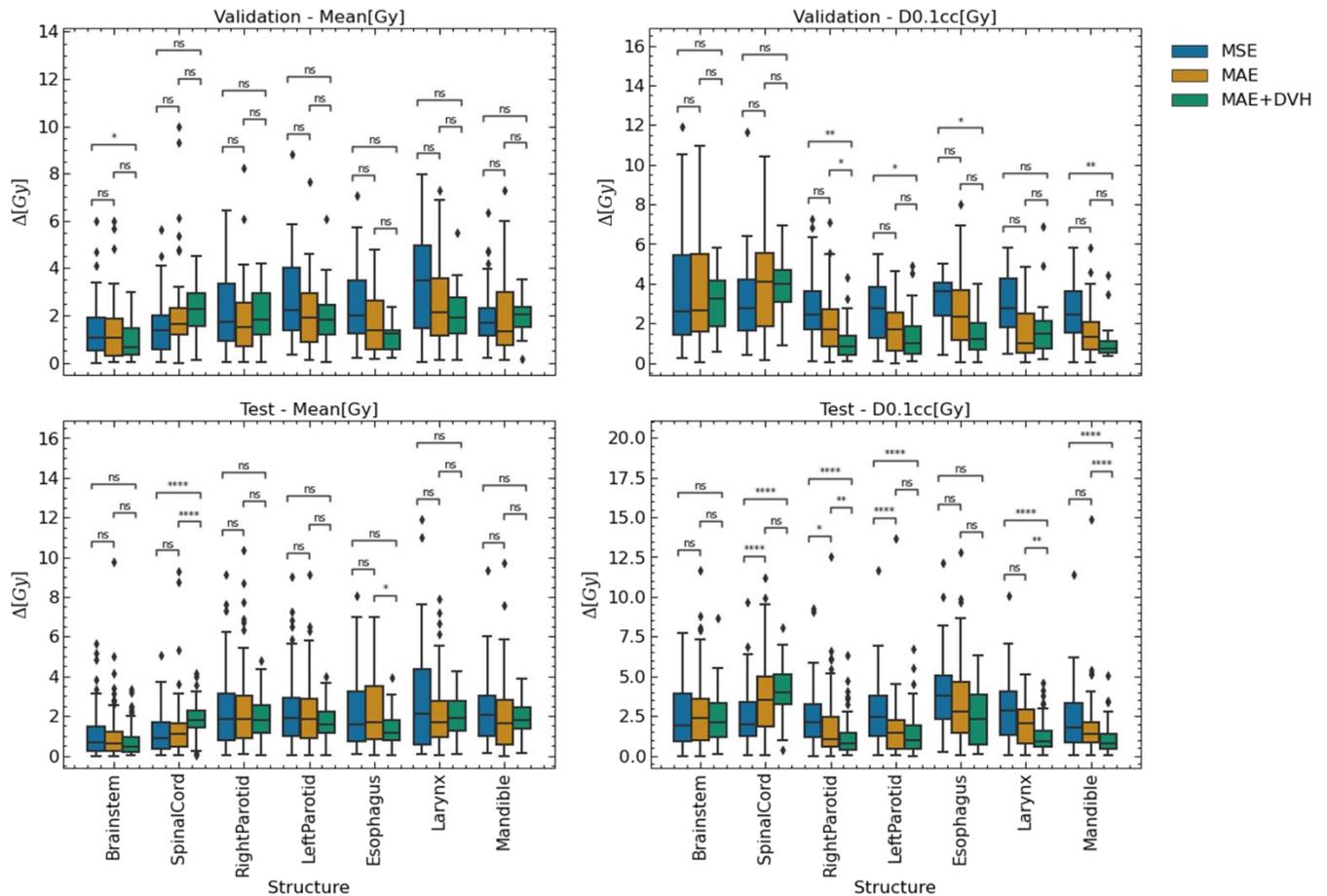

Figure 5 OAR DVH metrics (Mean[Gy] and D0.1cc[Gy]) comparison among MSE, MAE, MAE+DVH models on the validation and test datasets. Wilcoxon test was used for analysis. Star (*) values shows different $P$ values, i.e., $*P< 0.05$, $**P< 0.01$, $***P< 0.001$, $****P< 0.0001$, ns = no significance ($P>0.05$).

## Comparing Statistics of OARs DVH Metrics

 demonstrates no statistically significant difference among the three models on validation and test datasets in terms of Mean[Gy], while the MAE+DVH model significantly outperformed the MAE and MSE models in terms of D0.1cc[Gy] for the parotids and mandible for both the validation ($P < 0.01$) and test ($P < 0.0001$) datasets.

## Comparing Statistics of Targets DVH Metrics

Model performance for target coverage metrics (D1%[Gy], D95%[Gy] and D99%[Gy] for PTV56, PTV63, and PTV70) are intercompared in Figure 6. The MAE model performed significantly better than the MSE model on validation datasets. The initial comparison results between MSE and MAE, where MAE outperformed MSE, pointed towards adding the DVH loss to MAE instead of MSE. The MAE+DVH model significantly outperformed both MSE and MAE for target metrics on validation and test datasets (maximum *P*<0.05).

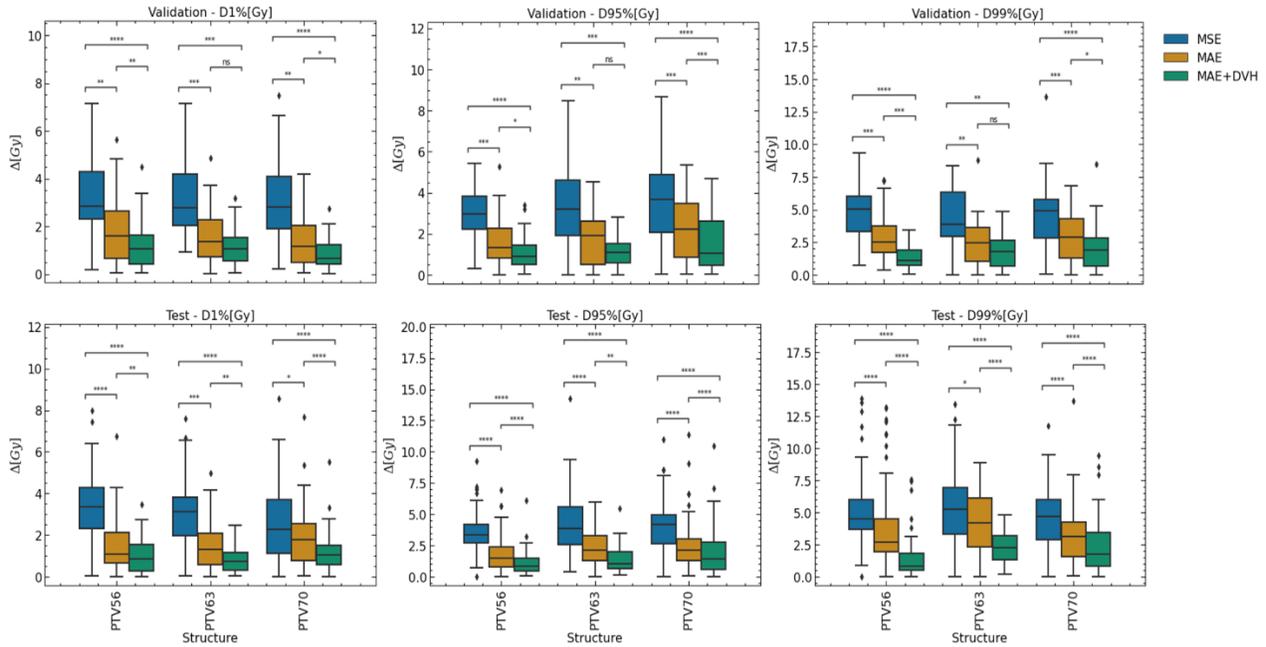

Figure 6 Targets DVH metrics (D99%[Gy], D95%[Gy], and D1%[Gy]) comparison among MSE, MAE, MAE+DVH models on the validation and test datasets. Wilcoxin test was used for analysis. Star (*) values shows different *P* values, i.e., **P*< 0.05, ***P*< 0.01, ****P*< 0.001, *****P*< 0.0001, ns = no significance.

Visual Comparison between Predicted and Reference Dose Distribution

Figure 7 compares slices of the reference and predicted dose and corresponding DVH curves for a selected test case with predicted 3D dose and DVH scores of 1.95Gy and 2.44Gy, respectively. High similarity between the predicted and reference DVH curves is observed.

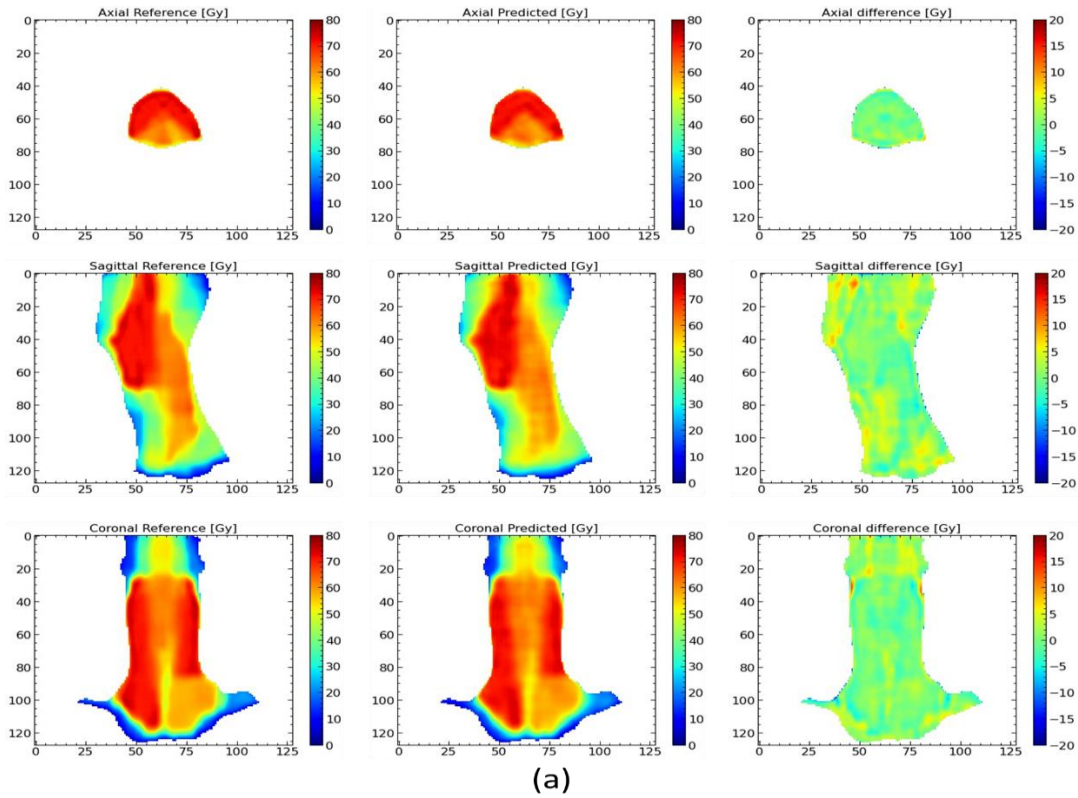

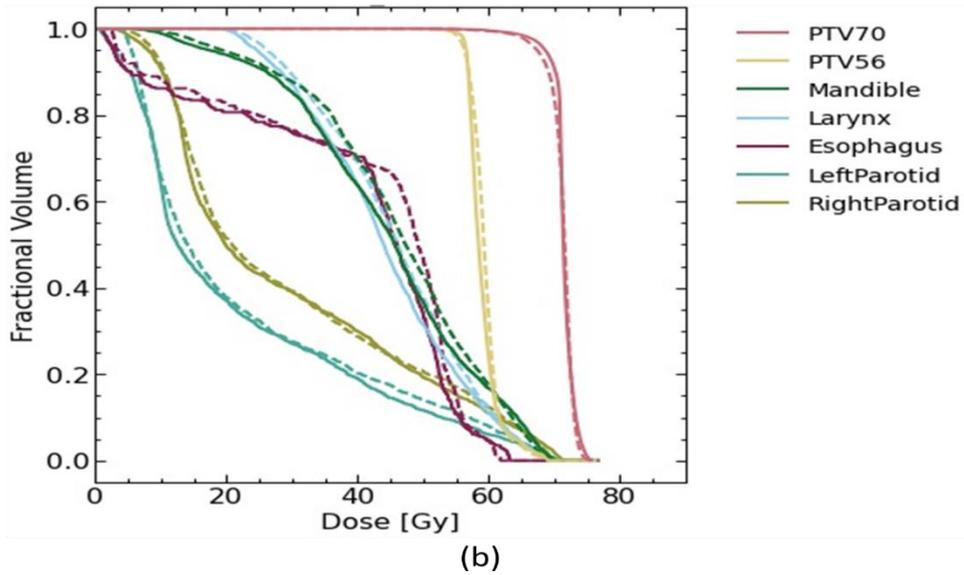

Figure 7: Comparison of reference and predicted dose distributions of two random test cases from test datasets were randomly selected for a demonstration. The reference dose distribution (first column), predicted dose distribution (second column), their voxel-wise differences (third column), and histograms of their voxel-wise differences (fourth column). The solid lines denote the reference DVH bends of targets and OARs, whereas dashed lines represent their corresponding predicted DVH bends. The solid lines and dashed lines are overlapped.

DeepDoseNet had the best performance on the OpenKBP dose metric. Direct comparison with other participants can be found on the OpenKBP site.

## Discussion

In this study, a compact 3D CNN, DeepDoseNet, is developed for predicting 3D dose distribution. The model utilizes ResNet and Dilated DenseNet for dense feature extraction. DeepDoseNet's performance varies with the training loss function. The customized MAE and MAE+DVH loss functions improved the dose prediction accuracy for PTVs and some OARs when compared to the standard MSE. When trained with MAE+DVH, DeepDoseNet predicts the 3D H&N dose distributions within 2 Gy. (Table 1– 3D dose score – test dataset).

DeepDoseNet MSE and MAE network entries have placed 23$^{rd}$ and 20$^{th}$ in the OpenKBP dose rankings, respectively [1]. The addition of the DVH metric is responsible for the performance improvement to the top ranked dose entry. The performance of the other OpenKBP participants, who used different network strategies typically with an MSE-based loss function, along with our findings for the performance with different loss functions, suggests that altering the training loss function can have as much of an impact on the performance as does the network design. Our MAE-based loss functions may have performed well, using the same mathematical form as the OpenKBP metrics used to rank participants. Other participants might improve their ranking by altering their training loss function.

On the OpenKBP DVH rankings, DeepDoseNet trained with MAE+DVH ranks at 18 amongst the participants with the test data sets[†]. The MSE and MAE models have ranked at 25$^{th}$ and 22$^{nd}$ among the participants. In comparing our training loss functions, the DVH tests score improved by a similar amount when going from MSE (3.0 Gy) to the MAE (2.3 Gy) as when going from MAE to MAE+DVH (1.6 Gy).

The use of the MAE+DVH loss function improved the model's prediction in almost every aspect. Interestingly, the improvement was greatest for targets, where the dose is homogeneous compared to the OARs, which contain steep dose gradients if they are not in low-dose regions. This aligns with the primary motivation for introducing MAE; the minimization of MSE in high dose gradient regions compromises the accuracy in homogeneous dose regions. An interesting exception to the MAE+DVH improvement is for the spinal cord, where MSE had the best performance in terms of the mean and max dose.

---

[1]OpenKBP continues to accept entries. Rankings as of 10/22/21
https://competitions.codalab.org/competitions/23428#results

Looking only at the DVH metric score, MAE+DVH improved all PTV metrics with statistical significance. OARs, however, had mixed results. A contributing factor to this may be the use of the down-sampled dose and structure masks. The reduction resulted in structure-mask and dose-voxel dimensions of ~3.5×3.5×2 mm$^3$, with a volume of 0.04 cc. At this resolution, the PTV DVH metrics (D99%[Gy], D95%[Gy], D1%[Gy]) are relatively stable with low variance since PTVs have relatively large volumes (100's of cc), and the desired PTV dose is homogeneous. However, for small volume OARs, the poorer structure mask resolution coupled with the high dose gradient fall-off outside the PTV can result in a large variance in DVH metric values. For all OARs, the D0.1cc[Gy] evaluation results in a ~2-3 voxel evaluation on the structure mask, which can have a large variance. Large DVH metric variation can lead to failure of the statistical tests to demonstrate significance among different loss functions when analyzing some OARs. The use of non-down-sampled structure masks might reveal statistical significance between MAE+DVH and the other methods.

Several OpenKBP participants have published their methods. Gronberg et al. [58] tested multiple network architectures. Their highest-performing model used a 3D Unet model, which incorporated a dilated DenseNet block between the Unet's encoder and decoder, a weighted MSE loss function, and a patch-based approach. Zimmermann et al.[59] utilized a 3D Unet with additional ResNet blocks in the Unet's encoder and decoder, an L1 (MAE) loss combined with a feature loss which uses a pre-trained video classifier as a feature extractor. Nguyen et al. [60] also used a 3D Unet–based architecture, along with Monte Carlo dropout and bootstrap aggregation to improve performance. Liu et al. [61] cascaded two 3D Unets. Our network shares features seen in these other models; Unet-based with ResNet blocks, a dilated DenseNet with convolution layers of different dilation rates, and an MAE loss function, although our combination is unique. Our custom MAE+DVH loss function is also unique, likely the differentiator that resulted in DeepDoseNet achieving the best-ranked dose score.

The limitations of this study are common for all OpenKBP participants. The first is data size. Deep learning-based methods often improve with larger datasets. Although DeepDoseNet did not exhibit any over-fitting during training as validation loss decreased alongside training loss, more data could tighten the training and validation loss curves distance. Second, OpenKBP only included one specific disease site (i.e., head and neck) with seven limited OARs, three prescribed targets and doses estimated by nine field IMRT. Changing sites, OARs, prescription doses, and delivery techniques will require remodeling. Nonetheless, DeepDoseNet should be broadly applicable to any disease site and delivery method without specific changes required in the current network architecture, except possibly the data arrangement as per the network's input/output requirements.

Although DeepDoseNet achieved a 2 Gy dose-score, its clinical utility is untested. DeepDoseNet (and the other OpenKBP participants) method(s) may be useful in KBP planning paradigms to predict achievable DVH metrics and the 3D dose distribution. However, simply reproducing the deep-learning-based dose prediction with a deliverable treatment plan does not necessarily result in a Pareto optimal or clinically usable plan. Like other planning methods, the final dose distributions must be carefully reviewed for clinical acceptability, with potential alterations to suit the specific patient case. Beyond clinical planning, DeepDoseNet, and other deep-learning-based dose predictions could be useful in other applications, such as predicting the priority of patient-specific OARs and their need for careful delineation review [62], thus improving efficiency in other aspects of the clinical workflow.

## Conclusions

DeepDoseNet trained with the MAE+DVH loss function predicts voxel-wise head-and-neck IMRT dose distributions within 2 Gy in just a few seconds on a modern GPU. The use of the DVH-based loss function reduced $\overline{S_D}$ by ~60% and $\overline{S_{DVH}}$ by ~70%, and resulted in the best performance amongst OpenKBP entries[†]. DeepDoseNet should be suitable for automated radiation therapy workflows, including clinical planning and studies of the dosimetric relevance of delineated organs-at-risk. DeepDoseNet mimics an optimization engine providing reasonable dose distributions, and in the future, it can be added to a pipeline to run *"in silico"* dosimetric studies.

## Acknowledgment

The NIH grant R01CA222216-01A1 supported this work